\documentclass[superscriptaddress,floats,aps,prl,preprint]{revtex4}
\usepackage{graphicx}
\usepackage{amsmath}
\usepackage{amssymb}
\usepackage{dcolumn}
\usepackage{color}
\begin{document}
\title{Quantum simulation of low-temperature metallic liquid hydrogen}
\author{Ji Chen}
\affiliation{ICQM and School of Physics, Peking~University,~Beijing~100871,~P.~R.~China}
\author{Xin-Zheng~Li}
\email{xzli@pku.edu.cn}
\affiliation{School of Physics, Peking~University,~Beijing~100871,~P.~R.~China}
\author{Qianfan Zhang}
\affiliation{ICQM and School of Physics, Peking~University,~Beijing~100871,~P.~R.~China}
\author{Matthew~I.~J.~Probert}
\affiliation{Department of Physics,  University of York, York YO10 5DD, U.K.}
\author{Chris~J.~Pickard}
\affiliation{Department of Physics and Astronomy, University College London, London WC1E 6BT, U.K.}
\author{Richard~J.~Needs}
\affiliation{Theory of Condensed Matter Group, Cavendish Laboratory, 
   University of Cambridge, J. J. Thomson Avenue, Cambridge CB3 0HE, U. K.}
\author{Angelos~Michaelides}
\affiliation{London Centre for Nanotechnology and Department of Chemistry, University College London, London WC1E 6BT, U.
K.}
\author{Enge Wang}
\email{egwang@pku.edu.cn}
\affiliation{ICQM and School of Physics, Peking~University,~Beijing~100871,~P.~R.~China}

\date{\today}

\begin{abstract}

  Experiments and computer simulations have shown that 
  the melting temperature of solid hydrogen drops with pressure above about 65 GPa,
  suggesting that a liquid state might exist at low
  temperatures.
  It has also been suggested that this low temperature liquid state
  might be non-molecular and metallic, although
  evidence for such behaviour is lacking.
  Here, we report results for hydrogen at high pressures using \textit{ab initio}
  path-integral molecular dynamics 
  methods, which include a description of the quantum motion of the protons
  at finite temperatures.
  We have determined the melting temperature as a function of pressure
  by direct simulation of the coexistence of the solid and liquid phases
  and have found an atomic solid phase from 500 to
  800~GPa which melts at $<$200 K. 
  Beyond this and up to pressures of 1,200~GPa a metallic atomic liquid is stable
  at temperatures as low as 50 K.
  The quantum motion of the protons is critical to the low melting temperature in 
  this system as \textit{ab initio} simulations with classical nuclei lead to a considerably higher
  melting temperature of $\sim$300~K across the entire pressure range considered.

\end{abstract}

\pacs{31.15.A-, 62.50.-p, 71.15.Pd, 81.30.-t}


\maketitle

\clearpage


Ever since Wigner and Huntington's prediction in 1935 that solid molecular hydrogen
would dissociate and form an atomic and metallic phase at high pressures~\cite{wigner},
the phase diagram of hydrogen has been the focus of intense experimental and 
theoretical studies.
Advanced experimental techniques, notably diamond anvil cell approaches, mean that it 
is now possible to explore hydrogen at pressures up to about
360 GPa~\cite{mao1994,loubeyre2002,eremets,zha}, and new types of diamond anvil cell may be able to 
access much higher pressures~\cite{dubrovinsky_2012}.
These experiments, along with numerous theoretical studies, have revealed a remarkably 
rich and interesting phase diagram comprising regions of stability for 
a molecular solid, a molecular liquid and an atomic liquid, and
within the solid region four distinct phases have been detected~\cite{howie,goncharov,eremets,zha}.
Furthermore, metallic hydrogen has been observed in high temperature
shock wave experiments~\cite{fortov,weir,nellis,hicks} and it is accepted to be a major component
of gas giant planets such as Jupiter and Saturn~\cite{exoplanet}.
Despite the tremendous and rapid progress, important gaps in our understanding of the phase diagram 
of high pressure hydrogen remain, with arguably the least well understood issue being the solid to liquid melting
transition at very high pressures.
Indeed the melting curve is only established experimentally and theoretically 
up to around 200 GPa~\cite{datchi,gregoryanz,deemyad,eremetsjetp,bonev2004}.
However, from 65 GPa up to about 200 GPa the slope of the melting 
line is negative (i.e., the melting point drops with increasing pressure), which suggests that
at yet higher pressures a low temperature liquid state of hydrogen might exist or, 
as suggested by Ashcroft, perhaps even a metallic liquid state at zero K~\cite{ashcroft2004}.
Further interest in hydrogen at pressures well above 200 GPa stems from other remarkable 
suggestions such as superfluidity~\cite{ashcroft2000jpcm} and superconductivity at room
temperature~\cite{ashcroft1968,gross2008,mcmahon},  
all of which imply that hydrogen at extreme pressures could be one of the most interesting 
and exotic materials in all of condensed matter.

In the current study we use computer simulation techniques to probe 
the low temperature phase diagram of hydrogen in the ultra-high 500-1,200 GPa regime
to try and find this potential low temperature liquid state of hydrogen.
We describe the proton motion using \textit{ab initio} path-integral
molecular dynamics (PIMD)
methods~\cite{marxpimd1,marxpimd2,tuckermanpimd1,tuckermanpimd2,qfzhang,xzliprl,xzlipnas},
which are based on forces computed ``on the fly'' as the dynamics of
the system evolves, and can account for bond formation and breaking in
a seamless manner.
To compute the melting line we simulate solid and liquid phases in
coexistence~\cite{morris,ogitsu,alfe,bonev2004}. 
The coexistence approach minimises hysteresis effects arising from super-heating or
super-cooling during the phase transition, and
although it has been used with model potentials and \textit{ab initio} approaches before, the
current study is, to the best of our knowledge, the first to combine
it with the \textit{ab initio} PIMD method. 
With this combination of 
approaches we have found a low-temperature metallic atomic liquid phase at
pressures of 900~GPa and above, down to the lowest temperature we can
simulate reliably of 50~K.
The existence of this low temperature metallic atomic liquid is
associated with a negative slope of the melting line between atomic
liquid and solid phases at pressures between 500 and 800~GPa.
This low temperature metallic atomic liquid is strongly quantum in
nature since treating the nuclei as classical particles significantly
raises the melting line of the atomic solid to $\sim$300~K over the
whole pressure range. 
It also destroys the negative slope of the melting line,
and consequently it does not predict a low temperature liquid phase.
Our study provides a finite temperature phase diagram of hydrogen
under ultrahigh pressures and gives direct numerical support for
Ashcroft's low temperature metallic liquid.

\begin{figure}[h]
\centering
\includegraphics[width=0.70\linewidth]{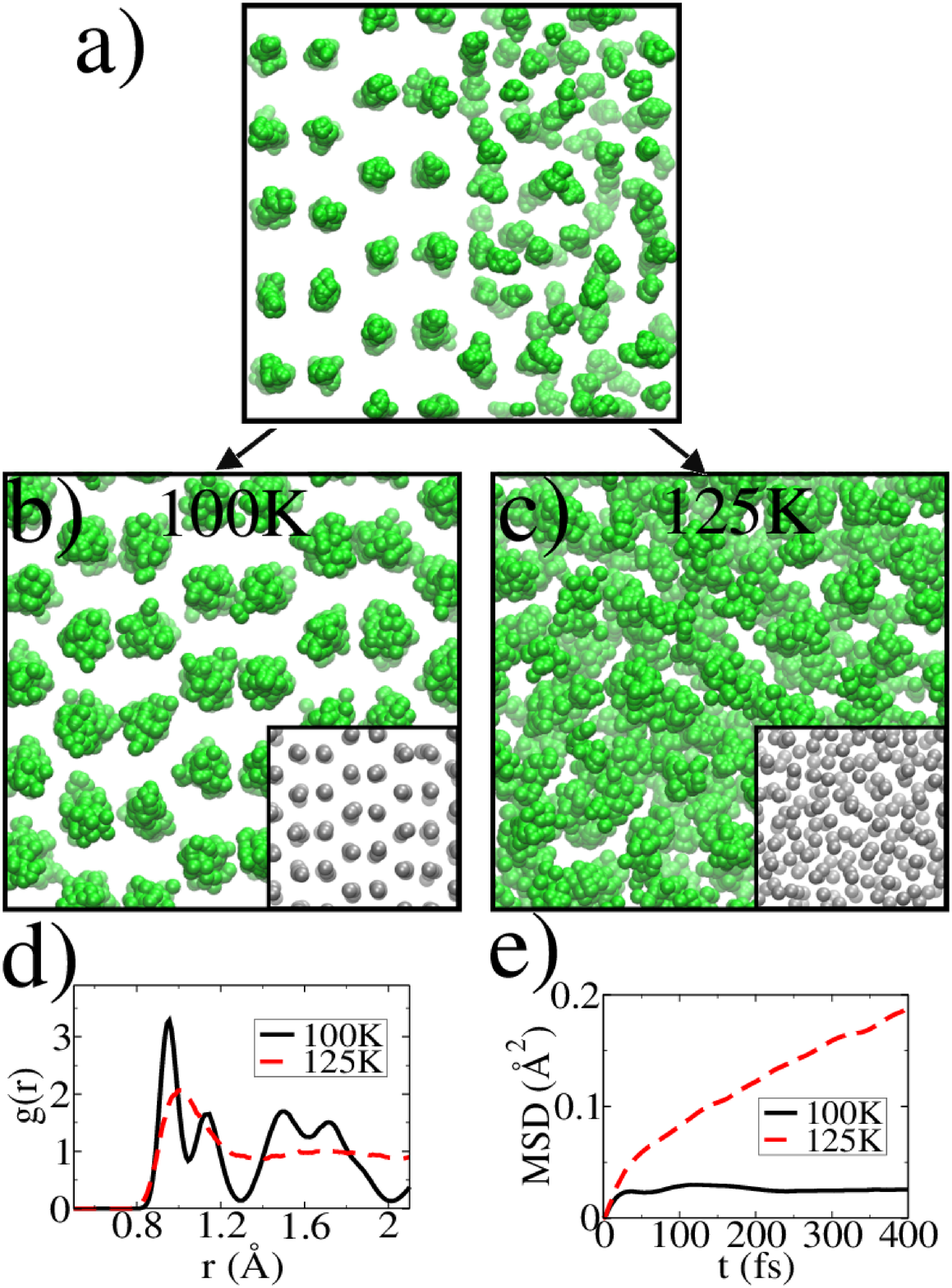}
\caption{\label{figure1}
\textit{Ab initio} PIMD simulations of solid-liquid coexistence and melting. Snapshots 
of the PIMD simulations at 700~GPa showing (a) the starting structure, 
(b) the final state at 100~K, and (c) the final state at 125~K.
32 beads~(green balls) were used to represent the imaginary time path-integral for each atom.
The grey balls in the insets of (b) and (c) correspond to the centroid of each atom.
(d) Radial distribution functions (RDFs) for the same two simulations at 100 and 125 K. 
At 100~K (black solid line) the solid state persists, whereas at 125~K (red dashed line)
the RDF is characteristic of a liquid. 
(e) Mean square displacements 
(MSD) as a function of time from separate adiabatic centroid MD simulations within
the path-integral framework.
%
The MSD for the 100~K solid phase saturates rapidly, whereas for the liquid phase at 125~K 
it rises approximately linearly with time, resulting in a finite diffusion
coefficient.  }
\end{figure}

Extensive computational searches for low enthalpy solid structures of
hydrogen have been performed using density functional theory (DFT)
methods~\cite{pickard2007,ceperley2011,pickard2012,pickard2012b,yanming} and recently a systematic
analysis of the evolution of the low-enthalpy phases has been provided
by Labet \textit{et al.}~\cite{labet1,labet3,labet4}.
These searches found a metallic phase of $I4_1/amd$ space group
symmetry to be stable from about 500 GPa to 1,200~GPa~\cite{ceperley2011},
when quasi-harmonic proton zero-point motion was included.
We use this phase, therefore, as the starting point for our finite temperature exploration
of the phase diagram and melting line. 
With the coexistence method we have performed a series of 
two phase solid-liquid simulations at different temperatures (from 50~K to 300~K) which 
are then used to bracket the melting 
temperature from above and below. 
We begin by considering the 500--800~GPa pressure regime and show an example of the data 
obtained from the coexistence simulations at 700~GPa in Fig.~\ref{figure1}.
At this pressure we find that for $T\geq$125~K the system transforms in to a liquid state, while for
$T\leq$100~K, it ends up as solid.
The phases on either side of the melting line (at this and other pressures) were characterized by 
their radial distribution function (RDF) and as can be seen in Fig.~\ref{figure1} (d) upon moving from
100 to 125 K the system clearly transforms from solid to liquid.
The phases were also characterised by the variations in mean
square displacement (MSD) of the nuclei of the particles over time.
Since PIMD rigorously provides only thermally averaged
information we used the adiabatic centroid molecular dynamics (MD) approach within the
path-integral scheme to obtain real-time quantum dynamical
information~\cite{adcmd}.  
Again, as shown in Fig.~\ref{figure1} (e), the distinction
between the solid phase at 100 K and liquid phase at 125 K is clear.

\begin{figure}[h]
\centering
\includegraphics[width=0.9\linewidth]{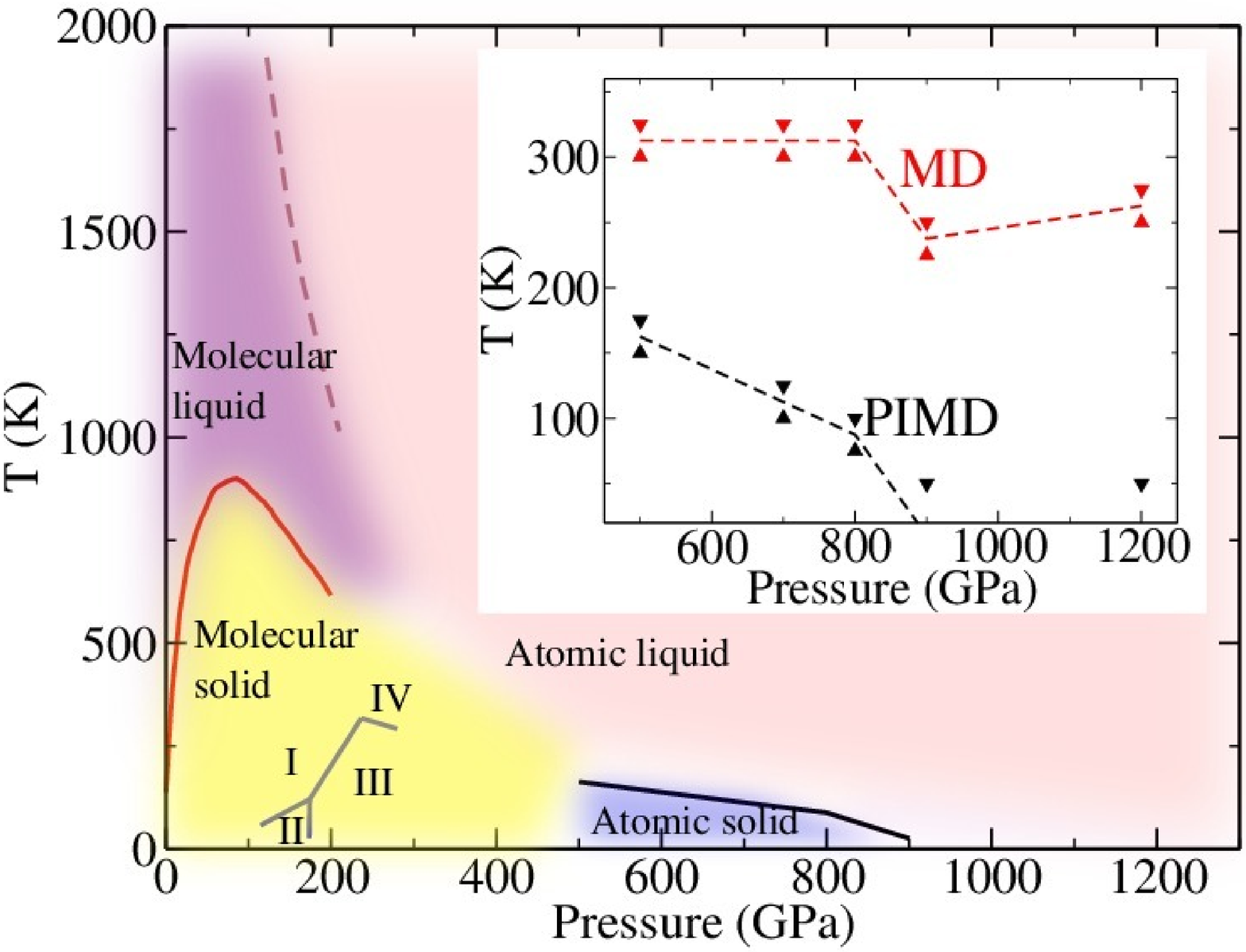}
\caption{\label{figure2}
Phase diagram of hydrogen with regions of stability for the molecular solid (yellow), 
molecular liquid (purple), atomic solid (blue), and atomic liquid (pink) indicated
by the various colours.
The dashed line separating the molecular and atomic liquid phases is taken from  
quantum Monte Carlo calculations~\cite{ceperley2010}.
%
The solid line separating the molecular solid and molecular liquid phases
is taken from \textit{ab initio} MD simulations~\cite{bonev2010}, whose negative
slope has been confirmed by several experiments~\cite{datchi,gregoryanz,deemyad,eremetsjetp}.
The thick black line is the melting curve obtained in this study from the
\textit{ab initio} PIMD coexistence simulations.
The solid lines separating phases I, II, III and IV are from Refs.~\cite{goncharov,howie}.
The inset shows how the high pressure melting curve (dashed lines) are established 
in our simulations.
The black and red triangles (inset) correspond to the PIMD and MD results, respectively.
The up triangles give the highest temperatures for solidification and
the down triangles show the lowest temperatures for liquefaction. }
\end{figure}

The same coexistence procedure was used to locate the melting point at 500 and 800 GPa, leading to the
melting line shown in Fig.~\ref{figure2}. 
The up~(down) triangles indicate the highest (lowest) temperatures at 
which the solid (liquid) phases are stable, bracketing the melting temperatures within a 25~K 
window. 
From this we see that the melting temperature is 150--175~K
at 500~GPa and that it drops rapidly with increasing pressure yielding a melting 
temperature of only 75--100~K at 800~GPa.
Thus the melting curve has a substantial negative slope
($dP$/$dT<0$) in this pressure range.
Across this entire pressure range the molten liquid state is atomic 
and the solid phase which grows is the original atomic $I4_1/amd$ phase that
was used as the starting structure.
Given that molecular phases have been observed at pressures lower than 
360~GPa in both experimental and theoretical studies~\cite{zha,loubeyre2002,kitamura},
we suggest that a molecular-to-atomic solid-solid phase transition should occur between
360 and the lowest pressure of 500~GPa considered here.

The negative slope of the melting line up to
800 GPa suggests that at even higher pressures a lower temperature liquid phase might
exist.
Motivated by this, we carried out simulations at 900 and 1,200~GPa.
However, in this pressure range we have to consider nuclear exchange
effects, which are neglected in the PIMD simulations performed here,
but which could potentially become significant.
Indeed, analysis of our simulations reveals that at these pressures the dispersion of the beads
in the path integral ring polymer becomes comparable to the smallest inter-atomic separations when the
temperature is below $\sim$40~K (see Fig.~S6). 
This is the so-called quantum degeneracy temperature,
below which the exchange of nuclei will be important.
With this in mind we performed all simulations in this very high pressure regime at 
$T \geq 50$~K.
Interestingly we find that at 50~K at both 900 and 1,200 GPa the systems are 
already in the liquid state, revealing that the melting temperate at these pressures
is $<$ 50 K. 
Whether the liquid phase is the 0~K ground state of hydrogen at these pressures is
not something we can establish at this stage.
However, the large negative slope of the melting line at lower pressures and
the observation of a liquid phase at temperatures as low as 50~K present strong support for
Ashcroft's low temperature liquid metallic state of hydrogen~\cite{ashcroft2004},
and it implies that any room temperature superconductor in this regime would have to
be a liquid.

It is instructive to compare the results of the \textit{ab initio} 
PIMD simulations with those obtained from the \textit{ab initio} MD
approach in which the nuclei are approximated by classical point-like particles. 
To this end we performed a second complete set of coexistence simulations with \textit{ab initio} MD across the
entire 500--1,200 GPa range. 
The \textit{ab initio} MD melting line is shown by the red data in the inset of Fig.~\ref{figure2}, 
where it can be seen that the melting temperatures obtained from the MD simulations are much higher than
those from the fully quantum PIMD simulations. 
The \textit{ab initio} MD melting temperature is well above 200 K
across the pressure range 500--1,200 GPa, and the slope of the melting
line is small.
These remarkably large differences between the \textit{ab initio} MD and PIMD melting lines 
clearly demonstrate that quantum nuclear effects play a crucial role in the stability
of the low temperature liquid.

In conclusion, our results are consistent with a melting curve with a negative slope
between an atomic solid and a low temperature (50~K or
below) metallic atomic phase.
A comparison of the MD and PIMD results shows that the quantum nature
of the nuclei is responsible for the large negative slope of the
melting line and the consequent existence of the low temperature metallic liquid. 
A natural conclusion from this is that the phase diagram for hydrogen in the pressure regime
considered should exhibit rather large isotope effects. 
It is not clear from our studies whether the liquid phase survives
down to 0~K, and this needs to be studied using other methods.
The occurrence of the $I4_1/amd$ solid above
about 500~GPa indicates that a molecular-to-atomic solid-solid phase
transition must occur at lower pressures, which could be accessible to
experiments in the near future.

\section*{Methods}

Density-functional theory~(DFT) calculations 
with the PBE exchange correlation functional 
were performed with the VASP code~\cite{pbe, VASP1, VASP2, VASP3, VASP4}.
Projector augmented wave potentials were used along with a 500~eV plane wave 
cutoff energy~\cite{paw}. 
A Monkhorst-Pack $k$-point mesh of
spacing $2\pi \times 0.05\text{\AA}^{-1}$ was used along with a unit cell
that typically contained 200 atoms.
Convergence tests with larger unit cells and higher plane wave cutoff energies are reported
in the SI.

MD and PIMD simulations were of the ``Born-Oppenheimer'' type within the NVT ensemble, 
using a Nos\'e-Hoover chain thermostat~\cite{nose-hoover}.
A 0.5~fs time step was used and all simulations were run for at least 20,000 steps (10~ps). 
The PIMD simulations made use of a recent implementation 
of this method in VASP.
32 beads were used to sample the imaginary-time path-integral at each temperature, 
as this was found to be the minimum number of beads needed to obtain reliable 
estimates of the melting temperature (SI).
When calculating real-time atomic diffusion
in the liquid phase using the adiabatic centroid MD
method~(Fig.~\ref{figure1} (e)), a smaller time step of 0.05~fs was used
after thermalization and 20,000 steps were used to characterize the atomic diffusion.
In the coexistence simulations, each run was initiated from a
structure comprising a solid $I4_1/amd$ phase and a liquid phase
obtained by selecting samples from earlier simulations at higher
temperatures.
%
%
To test the sensitivity of the results to the chosen starting structures we performed
simulations with several different uncorrelated liquid phase
structures, and found that the results were independent
of the starting liquid structure.
The system was normally found to liquefy or solidify after about
1--2~ps and equilibrium was reached soon afterwards~(Fig.~S5).
For more computational details, please read the supporting online
materials.

\begin{acknowledgements}
  J.C., X.Z.L., and E.W.\ are supported by NSFC. A.M.\ is supported by
  the European Research Council and the Royal Society through a Royal
  Society Wolfson Research Merit Award.
  We are grateful for computational resources provided by the
  supercomputer TianHe-1A in Tianjin, China.
\end{acknowledgements}

\end{document}